\documentclass[smallextended,natbib]{svjour3}\usepackage[]{graphicx}\usepackage[]{xcolor}
\makeatletter
\def\maxwidth{ %
  \ifdim\Gin@nat@width>\linewidth
    \linewidth
  \else
    \Gin@nat@width
  \fi
}
\makeatother

\definecolor{fgcolor}{rgb}{0.345, 0.345, 0.345}

\usepackage{framed}
\makeatletter
 {\par\unskip\endMakeFramed%
 \at@end@of@kframe}
\makeatother

\definecolor{shadecolor}{rgb}{.97, .97, .97}
\definecolor{messagecolor}{rgb}{0, 0, 0}
\definecolor{warningcolor}{rgb}{1, 0, 1}
\definecolor{errorcolor}{rgb}{1, 0, 0}
\usepackage{alltt}  
\smartqed  % flush right qed marks, e.g. at end of proof
\usepackage{graphicx}
\usepackage{array,multirow,graphicx}
\usepackage{amsmath}
\usepackage{graphicx}
\usepackage{natbib}
\usepackage{url} % not crucial - just used below for the URL 
\usepackage{multirow}
\usepackage{mathtools}
\usepackage{amsfonts}
\usepackage{rotating}
\usepackage{booktabs}
\usepackage{verbatim}
\usepackage{mathrsfs}
\usepackage{hyperref}
\usepackage{pdflscape}
\usepackage{caption}
\usepackage{tablefootnote}
\usepackage{verbatim}
\usepackage{setspace}
\usepackage[utf8]{inputenc}
\usepackage{authblk}
\usepackage{pdfpages}
\usepackage{fancyhdr}

\IfFileExists{upquote.sty}{\usepackage{upquote}}{}
\begin{document}

\title{SpICE: An interpretable method for spatial data}

\author{Natalia da Silva \and
        Ignacio Alvarez-Castro\and
        Leonardo Moreno\and
        Andrés Sosa}


\institute{Corresponding author: Natalia da Silva \at
              Instituto de Estadística (IESTA), Universidad de la República, Montevideo, Uruguay.\\
              \email{natalia.dasilva@fcea.edu.uy}\\
              ORCID:0000-0002-6031-7451\\
           \and
          Ignacio Alvarez-Castro \at 
          Instituto de Estadística (IESTA), Universidad de la República, Montevideo, Uruguay.\\
          \email{ignacio.alvarez@fcea.edu.uy}\\
          ORCID:0000-0003-1633-2432\\
       \and
          Leonardo Moreno \at
          Instituto de Estadística (IESTA), Universidad de la República, Montevideo, Uruguay.\\
          \email{leonardo.moreno@fcea.edu.uy}\\
          ORCID:0000-0003-1630-1361\\
             \and
             Andrés Sosa \at
          Instituto de Estadística (IESTA), Universidad de la República, Montevideo, Uruguay.\\
          \email{andres.sosa@fcea.edu.uy}\\
          ORCID:0000-0002-6007-4373\\
}

\date{Received:  / Accepted: }

\maketitle

\begin{abstract}
Statistical learning methods are widely utilized in tackling complex problems due to their flexibility, good predictive performance and its ability to capture complex relationships among variables. Additionally, recently developed automatic workflows have provided a standardized approach to implementing statistical learning methods across various applications. However these tools highlight a main drawbacks of statistical learning: its lack of interpretation in their results.
In the past few years an important amount of research has been focused on methods for interpreting black box models.  Having interpretable statistical learning methods is relevant to have a deeper understanding of the model. In problems were spatial information is relevant, combined interpretable methods with spatial data can help to get better understanding of the problem and interpretation of the results.

This paper is focused in the individual conditional expectation (ICE-plot), a model agnostic methods for interpreting statistical learning models and combined them with spatial information. ICE-plot extension is proposed where spatial information is used as restriction to define Spatial ICE curves (SpICE). Spatial ICE curves are estimated using real data in the context of an economic problem concerning property valuation in Montevideo, Uruguay. Understanding the key factors that influence property valuation is essential for decision-making, and spatial data plays a relevant role in this regard.

\end{abstract}

\newpage

\section{Introduction \label{intro}}

Statistical learning methods have been used with high success in many fields and different kind of complex research problems. Some of the reasons of statistical learning method extended use is due to its flexibility, good predictive performance and its ability to capture complex relationships among variables. In recent years several computational tools to automate the workflow of statistical learning models implementation has been proposed, such as: caret \citep{caret}, h2o \citep{h2o}, tidymodels \citep{tidymodels} among others. These tools allows to train and tune many different algorithms, and choose the best one based on a selected performance measure in a standardized way reducing some common implementation errors. At the same time the automatic workflows results in models usually shows very good predictive performance but are hard to explain its decisions. These tools makes more evident the inability to explain key aspects of the problem under consideration in statistical learning methods. Fortunatelly, there is a growing amount of research focused on methods for interpreting black box models. Broadly speaking, \textit{interpretability} is the degree a human can understand decisions or predictions from a statistical method \citep{miller2019explanation}. Having interpretable methods can be useful to detect bias, understand model errors, improve the model performance, understand hidden relationships discovered by the algorithm among others. These reasons are relevant even when the objective is purely predictive. 

Interpretable methods can be divided, model specific or model agnostic. Model specific refers to models that are intrinsically interpretable based on model characteristics, for example regression coefficients. In this case since the interpretation is intrinsic to the model can be difficult for model comparison. On the other hand model agnostic are methods used after model fitting applying techniques that allow to analyze the results of any subsequent model to be trained. A revision of the different methods for interpretability can be found in \cite{molnar2020}, also \cite{maksymiuk2020landscape} presents a landscape of computational options and a complete taxonomy of the interpretable methods available in R contributed packages. The present work, is focused on model agnostic methods, mainly individual conditional expectation (ICE-plot), proposed by \cite{goldstein2015peeking}. ICE curves are used for visualize the relationship among response variable and a specific feature for each individual observation. ICE-plot might help to represent heterogeneous effects in a problem but might difficult to work in big data scenarios.  

In problems where data contains spatial structure, it could be use to improve the interpretability of statistical learning models in a natural way. Spatial data consist in any type of data which references a specific geographical area or location. In this context, different statistical methodologies have been proposed to perform clustering in spatial applications (see \cite{grubesic2014,varghese2013} for reviews). In \cite{chavent2018}, a hierarchical clustering algorithm with spatial restrictions is proposed, where the goal is to obtain homogeneous clusters with spatial contiguity. 

This paper explores some of the most used model agnostic interpretable methods that can be applied to any supervised statistical learning method in real state market application were the spatial information is relevant. The main goal of this paper is to combined spatial data with interpretable statistical learning methods. An extension of ICE-plot is proposed which takes into account spatial restrictions to group the ICE curves. Spatial ICE curves allows an easier interpretation in problems where the number of observations is big and the spatial information is relevant.

The motivating example we use in the paper is an application of statistical learning methods in property valuation. Real estate market has a key role in the economic activity and  have a central role in economic and financial crisis \citep{mooya2016standard}. Five statistical learning methods are used as predictive models of asking prices apartments in Montevideo, that were selected based on predictive performance measures. %The global financial crisis of 2007-08 pointed out how real state market can create massive financial instability. 

This financial market differs from other  markets due to the property value heterogeneity. Understanding of spatial variability of property prices is a relevant economic problem for several financial and government organizations. An adequate model for property valuation is an important tool in the decision making process in public and private sectors \citep{osland2010,case2004}. Several statistical methods have been used to identify significant patterns in home pricing, such as traditional hedonic models  \citep{rosen1974} to advanced in methods including spatial dependency and non-linear relationships. Usually these models consider building characteristics information such as location, structural characteristics and some variables related with neighborhood information to determine the price \citep{dubin1988estimation}. 

Many proposed models assume that property price can be decompose linearly as the sum of its determinants. These models are easy to explain and to obtain the joint variables impact on price. However, are also often not as good with respect to predictive performance, because the relationships that can be learned are very restricted and generally oversimplify reality. For these reasons, statistical learning methods and non-linear models  have grown significantly in the last years in real industry  \citep{limsombunchai2004, yoo2012,  park2015, goyeneche2017}. 

Currently, several supervised, unsupervised and semi-supervised learning methods are applied to large databases of properties with information on prices, asset characteristics and geographic attributes.  Although statistical learning methods have proven useful in real estate modeling for predictive purposes, the lack of transparency and intrepretation limits their use.

Two groups of explanatory factors were considered. In the first place, spatial information relative to the location based on geographical coordinates of the real estate. These variables are known to be key determinant of a real estate value \citep{kiel2008}. The neighborhood, general service access, the value of nearby properties, distance to special points of interest (downtown, a coast, etc), crime rate, are some examples of factor that have a large impact on property price. In second place, a complementary set of explanatory variables are features of the real estate itself, such as the number of bedrooms, number of bathroom,  total area of property, these are known in the property literature as hedonic variables \citep{sirmans2005}. 

The paper is structured as follows. Section \ref{section:2} presents an introduction to interpretable machine learning methods, along with the inclusion of the so-called Spatial ICE curves. Section \ref{section:3},  presents the results obtained. The section starts by providing a description of the real estate data in Montevideo (Uruguay) that was used for the study. Using this data, various statistical learning models were trained, and the main interpretable methods were applied. Finally, the results were compared with the proposed Spatial ICE curves (SpICE). Section \ref{section:4} concludes by discussing some final remarks and suggesting potential avenues for future research.

\section{ICE curves in spatial problems} \label{section:2}
Consider a supervised problem where the objective of statistical learning models is to approximate 
\[
\mathbb{E}(Y|X=x) = f(x) \approx \hat f(x);
\]
where $X=(X_1,X_2,\dots, X_q)$ is a vector of $q$-variables, $Y$ is the response variable and $\hat f$ is the fitted model that predicts the scalar $Y$ as a function of $X$. In this context, a goal of interpretable methods is to characterize the dependence of the ``main effects'' on $f(x)$ in each explanatory variable. It is also possible to analyze the ``low order'' dependence between pairs of variables. 

Lets assume the main goal is to understand the effect of a set of explanatory variables denoted as $X_S$ on the response variable in a model agnostic way ($S \in \{1,2,\dots,q\}$ and denote the subset $C$ as the complement of $S$). One of the earliest methods for this is partial dependence plot \citep{friedman2001}. The PD-plot describes the change of the response variable in a model as a function of the marginal effect of one or more variables (subset $X_S$) when averaging the effects of the other explanatory variables (complement subset $X_C$). 

Main advantages of PD-plot is that its estimation is very intuitive and presents a causal interpretation in the results of any learning model. The main drawbacks are that hide heterogeneous effects, it might be rely on unrealistic set of observations and is computationally expensive. Alternatives to PD-plot has being proposed to overcome its problems, such as accumulated local effects (ALE-plot) \citep{apley2020visualizing} and the individual conditional expectation (ICE-plot) \citep{goldstein2015peeking}.

\subsection{Individual conditional expectation (ICE-plot)} \label{sub:ICE}

The ICE-plot is proposed as an extension of PD-plot, to visualize the dependence of the prediction on a feature for each sample separately, with one curve per observation. The method attempts to capture the dependency of the response variable on a set of variables, allowing heterogeneous effects.

For each observation, the curve $\hat f_{S,ICE}^{(i)}(x_S)$ is obtained by varying $x_S$ in the function $\hat f(x_S,x_C^{(i)})$ while the variables $x_C^{(i)}$ remain constant. That is to say
\[
 \hat f_{S,ICE}^{(i)}(x_S)= \hat f(x_S, x_C^{(i)}). 
\]

It is worth noting that averaging the ICE curves corresponds the definition of the PD-plot
\[
\hat f_{S,PD}(x_S) = \frac{1}{N}\sum_{i = 1}^ N \hat f(x_s, x_C^{(i)});
\]
then is possible to interpret the ICE curves similarly as the classic PD-plot but allowing to pick interactions when visualizing the $N$ plots (something that vanish when all ICE are averaged into one curve).

One disadvantage of ICE plots is the overplotting when big data set are involved. In such cases, it can be challenging to discern any meaningful information due to the overlapping of data points. This is specially relevant for ICE curves since its main purpose is to look for heterogeneity patterns in the predictor effect. Usual graphical solutions such as transparency or 2-dimensional histograms are not suited for plotting lines.  Specifically for ICE-curves, \texttt{h2o} implementation (\cite{hall2017}) choose ICE-curves for decile values (in the observed response), so no matter how large is the sample data, only ten ICE curves are displayed, resulting in an oversimplified plot. An alternative is to plot a stratified sample of curves. This method can be combined with line transparency to allow to reduce the overplotting while is possible to see different patterns of ICE curves in the data. An application of this plot can be found in Figure \ref{fig-icesup} in  Section \ref{section:4}.

\subsection{ Spatial ICE curves (SpICE)} 
\label{seccD}
Spatial data is any type of data which references a specific geographical area or location. Each curve in the ICE-plot is associated with a specific observation, so it can be link to a geographical location in spatial applications. Combining ICE-plot and spatial structure of the data is possible to overcome some drawbacks of ICE curves and improve interpretability at the same time. 
In this paper,  SpICE  refers to clustering ICE curves with spatial contiguity constrains.

The Ward-like method is used for the construction of the clusters, see \cite{chavent2018}. This is based on two dissimilarity matrices ($D_0$ and $D_1$) and a mixing parameter $\alpha \in [0,1]$. $D_0$ is a dissimilarity matrix between ICE curves and $D_1$ is a dissimilarity matrix that consider the spatial problem.

Matrix $D_0$ is determined only based on a dissimilarity measure between each ICE curve. The key idea is to group together observations with similar predictor effect on the response that could be summarize as the level and change range of the ICE curve. For this reasons it is appropriate to consider a distance  that not only considers the value of the integral but also their growth/decrease. However, some ICE curves might not be differentiable in some point for a lot of statistical learning models. One way to overcome this issue and obtain estimates in a $C^1$ space, is to consider the curve $\widetilde{f}_{S,ICE}^{(i)}(x_S)$ that is the convolution of the function  $\hat{f}(x_S, x_C^{(i)})$  with a Gaussian kernel. That is,  
\begin{equation}\label{convo}
 \widetilde{f}_{S,ICE}^{(i)}(x_S) := \ \hat{f}(x_S, x_C^{(i)}) \ast K_h(x_S),
\end{equation}
where $K_h$ is the Gaussian kernel, $h$ the smoothing parameter and $\ast$ the convolution operation.  For the function $\widetilde{f}_{S,ICE}^{(i)}(x_S)$ in the equation (\ref{convo}) it is possible to use the Sobolev $W^{1,2}(\mathbb{R})$ metric \citep{sobolev},  induced by the norm
\[
\Vert  \widetilde{f}_{S,ICE}^{(i)}(x_S)   \Vert= \sqrt {\int \vert  \widetilde{f}_{S,ICE}^{(i)}(x_S)  \vert ^2 dt + \int \vert  \widetilde{f}_{S,ICE}^{\ '(i)}(x_S) (t) \vert ^2 dt}.
\]
Assuming the true $f(x)$ is a smooth function, the above transformation of ICE curve (the convolution) will not distort the ICE estimator in a relevant way. 

For the definition of the distance $D_1$  is necessary use the spatial information associated with each observation. Depending on the quality and nature of the spatial data available (such as coordinates points, or area data, etc) this metric distance could be defined in different ways. In the motivating example of this paper %with the data described in Section \ref{section:3}, 
the Euclidean distance between coordinates observations is used for simplicity.   

Let us consider a partition $\mathcal{P}_K = (C_1 ,\ldots , C_K )$ in $K$ clusters of ICE curves. Clustering algorithm proposed in \cite{chavent2018} is outlined in order to described how to choose a value for mixing parameter. First, consider $n$ observations, where $w_i$ represents the weight associated with the $i$-th observation. Let $D_0=[d_{0,ij}]$ and $D_1=[d_{1,ij}]$ be two dissimilarity matrices of size $n \times n$, associated with the $n$ observations, corresponding to the covariates and spatial constraints, respectively. Both dissimilarities are normalized to make them comparable. The pseudo-inertia of a group is now also indexed by the mixing parameter $\alpha$, which weighs the importance of the two dissimilarities. The mixed pseudo-inertia of the $k$-th group is defined as

$$I_{\alpha} \left( C^{\alpha} _k \right)=  (1-\alpha) \sum_{i \in C^{\alpha}_k} \sum_{j \in C^{\alpha}_k} \frac{w_i w_j}{2 \mu^{\alpha}_k}d^2_{0,ij}+\alpha \sum_{i \in C^{\alpha}_k} \sum_{j \in C^{\alpha}_k} \frac{w_i w_j}{2 \mu^{\alpha}_k}d^2_{1,ij};$$
with $\mu^{\alpha}k= \sum{i \in C^{\alpha}_k} w_i$, and the inertia of the partition with $K$ groups is given by
$$W _{\alpha}\left( \mathcal{P}^{\alpha}_{K} \right)= \sum_{k=1}^K I_{\alpha} \left( C^{\alpha}_k \right).$$
Secondly, given a partition into $K+1$ groups, now the Ward type algorithm is applied to obtain a new partition into $K$ groups, which is a solution to the following optimization problem,
$$\textrm{arg}  \min\limits_{A,B \in \mathcal{P}_{K+1}^{\alpha}}\, I_{\alpha}\left( A \cup B\right) - I_{\alpha}\left( A \right) - 
I_{\alpha}\left(B \right) .$$
Finally, a criteria are developed for choosing the mixing parameter $\alpha$ is based on the notion of the proportion of total mixed pseudo-inertia explained by the partition $\mathcal{P}_K^{\alpha}$ into $K$ groups,
$$Q_{\beta}( \mathcal{P}_K^{\alpha})= 1- \frac{W_{\beta} \left( \mathcal{P}_{K}^{\alpha} \right)}{W_{\beta}\left( \mathcal{P}_{1} \right)} \in [0,1]\,; \:\: \text{ with } \beta \in [0,1] \,,$$
where $W_{\beta}\left( \mathcal{P}_{1}\right)$ represents the total pseudo-inertia based on the dissimilarity matrix $D_0$ if $\beta=0$, the total pseudo-inertia based on the dissimilarity matrix $D_1$ if $\beta=1$, and if $\beta \in (0,1)$ it is the total mixed pseudo-inertia

The parameter $\alpha$ regulates the two factor's weight  involved in the algorithm. On the one hand, when $\alpha=0$ the clusters are just determined by covariates without spatial information and on the other hand when $\alpha=1$ the clusters are just determined by the spatial information. Ideally is desirable to determine $\alpha$ value which increases cluster's spatial continuity without loosing cluster homogeneity in terms of ICE curves.

\section{Application results}
\label{section:3}
%In this Section data description, predictive models with `h2o` and SpICE curves results are presented.
This Section start by describing the dataset used as motivating example, then performance results of statistical learning models and interpretable methods are presented. Finally, the SpICE curves results in the motivating example data are presented. 

\subsection{ Dataset: Appartment prices in Montevideo }
The data in this paper are from an eCommerce platform \href{https://www.mercadolibre.com.uy)}{Mercado Libre} where real estate are offered for sale and rental. The complete data set contains asking price information for properties in Montevideo, capital city of Uruguay from February 2018 to January 2019 \citep{Picardo}. There are 92,832 observations and 116 variables in the complete data set. Nevertheless, only apartments data will be consider for the analysis working with 70,817 apartment observations for sale in Montevideo.

\begin{figure}[hbpt]
\centering
\includegraphics[width=1\linewidth]{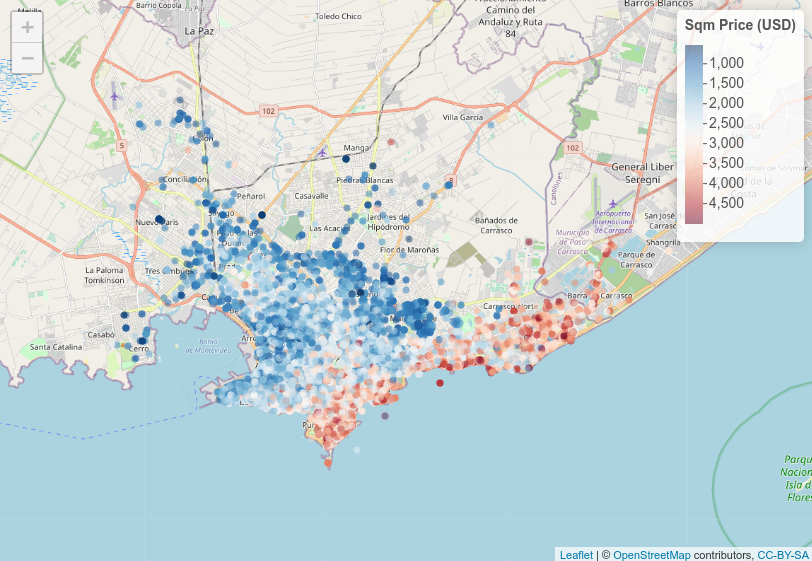}
\caption{Distribution of asking price apartments in Montevideo. Each dot represents one apartment and the color is the asking price per square meter in US dollars. 
\label{mapa} }
\end{figure}

 Figure \ref{mapa} shows the distribution of asking prices for apartments in Montevideo. Each dot on the plot represents one apartment available for sale, while the color of the dot indicates the asking price per square meter in US dollars.
Figure \ref{mapa} suggest locations close to the coast are associated with higher price range. Apartment's asking price ranges from  541 to 5000 US-dollars per square meters in Montevideo and the median value is close to 2500 US-dollars. 

\begin{table}[hbpt]
	\centering
	\caption{Variable description\label{tabvar} }
\scriptsize
\begin{tabular}{p{2.4cm}p{2.4cm}p{7cm}}
		\hline \\[-1.8ex] 
\textbf{Variable} & \textbf{Role} &  	\textbf{Description} \\
		\hline \\[-1.8ex] 	\hline \\[-1.8ex] 

\texttt{lpreciom2} & Response & Log of asking price in US-dollars per square meters. \\ \hline
  
\texttt{amenities} & Explanatory & Total number of amenities presented in the apartment.\\
\texttt{bedrooms} & Explanatory & Bedrooms quantity. Reduced to values between 0 (studio) and 3. \\
\texttt{bathroom} & Explanatory & Bathrooms quantity. Reduced to values between 1 and 3.\\
\texttt{elevators} & Explanatory & Elevators quantity. Reduced to values between 0 and 2. \\
\texttt{condition} & Explanatory & Property condition (new/used).  Properties with less than 1 year were marked as ``new''.\\
\texttt{expenses} & Explanatory &  Numerical value representing monthly expenses in Uruguayan pesos (local currency)\\
\texttt{garage} & Explanatory & Whether or not there is a garage. Reduced to values between 0 (``No'') and 1 (``Si'').\\ 
\texttt{ldistance\_beach} & Explanatory & Minimum distance (Euclidean)  between  the property   and the beach (in log scale). \\
\texttt{lsup\_constru} & Explanatory & Log of  area in square meters. Values over 2000  or under 9 square meter were remove from data. \\
\texttt{neighborhoodgr} & Explanatory &  Montevideo neighborhoods grouped by proximity in 12 regions. \\ \hline
\texttt{lat} & Spatial & Property latitude coordinate \\
\texttt{long} & Spatial & Property longitude coordinate \\
	\hline \\[-1.8ex] 	\hline \\[-1.8ex]  
	\end{tabular}
\end{table}
Table \ref{tabvar} presents the selected variables from the entire dataset to use for the statistical methods applied latter, indicating variable name in the dataset, its role in the statistical analysis and a brief description. All the models are fitted using as response variable the natural logarithm of the  price per square meters (\texttt{lprecio2}). The explanatory variables represent apartment features commonly used in real estate modeling, such as location (\texttt{neighborhoodgr}), total area, number of bedrooms and number of bathrooms. In addition, with the available information we create additional explanatory variables such as distance to the beach and total amenities. A reduced number of variables were selected to assure data quality from the complete data set. Finally, geographical coordinates of each property are used as additional information to improve interpretability using SpICE curves described earlier.

\subsection{Predictive models with \texttt{h2o} }
\label{modelo} 

Several models were trained %to estimate $f(x_i)$ 
using the automatic machine learning (\texttt{autoML}) procedure from \texttt{h2o} R package \citep{h2o} to predict the apartment  price as an alternative to classical methods. The  \texttt{autoML}  estimates models well tuned in 4 families: penalized linear models (glm), random forest (drf), xgboost, and Deep Learning. In the rest of the paper, the best model of each family and an additional metalearner are used. 

Table \ref{comparo} shows the  performance measures for the selected predictive models. The root mean square error (RMSE) and the $R^2$ are used to evaluate model performance. These values are computed with the response variable in logs (as it was used for trained every model). Additionally, mean absolute error (MAEo) and mean absolute percentage error (MAPEo) are both computed in the original scale of the response variable so can be read in dollars per square meter.  The four measures are computed using a testing data set different from training sample (2/3 training and 1/3 testing). 

\begin{table}[ht]
\centering
\caption{Predictive performance measures by model} 
\label{comparo}
\begin{tabular}{rlrrrr}
  \hline
 & modelo & rMSE & R2 & MAEo & MAPEo \\ 
  \hline
1 & stackedensemble & 0.14 & 0.82 & 245.27 & 9.68 \\ 
  2 & drf & 0.14 & 0.81 & 251.89 & 9.97 \\ 
  3 & xgboost & 0.15 & 0.80 & 265.28 & 10.46 \\ 
  4 & deeplearning & 0.19 & 0.65 & 372.30 & 14.68 \\ 
  5 & glm & 0.22 & 0.55 & 429.53 & 17.19 \\ 
   \hline
\end{tabular}
\end{table}

In terms of predictive performance, stacked metalearner, Random Forest and the XGboost algorithms show similar performance, somewhat better than the deep learning method or the penalized linear model. It is worth to note the stacked model combines 7 tree based models in its construction. On average, the best model get an average error in the asking price of \$245 per square meter, around 10\% of the observed price.  %The variability explained by  every model is between $0.74$ and $0.90$. 

\subsection{ Variable importance measures}
\label{4.2} 
The first approach to interpreting model results for statistical learning methods is to compute variable importance measures. Importance variable measures are  widely used for this problem. Some variable importance  measures are model specific but other are model agnostic and can be computed for any model. Explanatory variables with bigger values in this measure indicate that are more important for the predictive model and have more impact on the response variable. However, the weakness of this method is that has no information about the effect direction or shape.

\begin{figure}[htpb]
    \centering
    \includegraphics[trim=0cm 4cm 0cm 4cm, clip, scale=1]{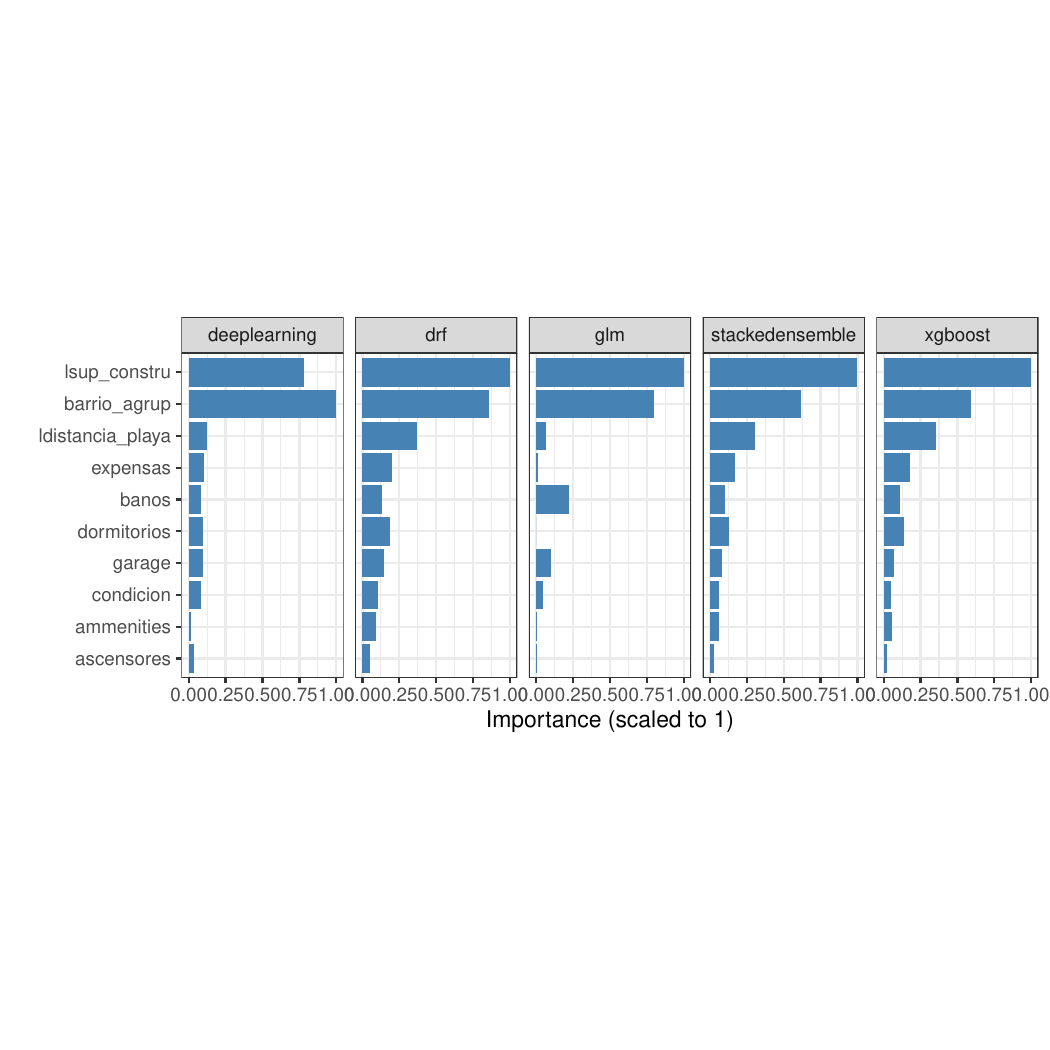}
    \caption{Variable Importance. Each panel represents a model, and the y-axis shows the variables included in all the models. Bar length represents the scaled variable importance measure to predict the apartment price.}
    \label{fig-imp}
\end{figure}

Variable importance measures were computed for each model. The measures were scaled so that the most important variable in each model has a value of 1, which simplifies the model comparison. 
Figure \ref{fig-imp} shows the results  by model for the variable importance. The predictors ordering are similar in all models.  The apartment area (\texttt{lsup\_constru}) and the neighborhood (\texttt{neighborhoodgr}) are the two most important variables to predict the apartment price. All the tree based methods (drf, xgboost and stackedemsemble) shows a third relevant variable which is the distance from the apartment to the beach (\texttt{ldistance\_beach}). The rest of the predictor variables are not relevant for prediction. 

\subsection{ Partial effect of apartment area } \label{sub:resultados_pdp}
Variable importance measure gives a ranking  of variables according to their relevance to predict the response variable.  However, they do not provide information on the effect that the individual variables have on the response. In order to characterize the average effect of apartment features on its  price, the PD-plot and the ALE-plot  are estimated. The algorithms used for its estimation  were done using R packages \texttt{pdp} \citep{pdp} and \texttt{ALEPlot} \citep{aleplot}. 
%interpretable methods described in Section \ref{section:2} are computed. Computations of the PD-plot, ALE-plot and ICE curves are done using R packages \texttt{iml} \citep{iml} and \texttt{ALEPlot} \citep{aleplot}. 
\begin{figure}[hbpt]
    \centering
    \includegraphics[trim=0cm 6cm 0cm 6cm, clip, scale=1]{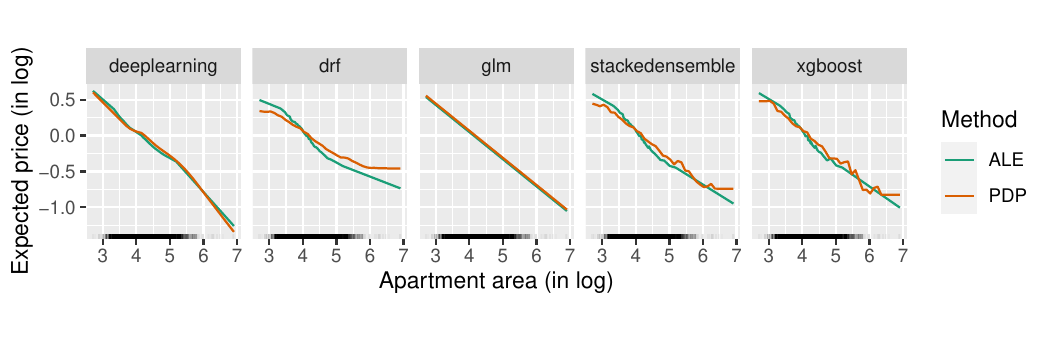}
    \caption{Effect of \texttt{lsup\_constru} variable in different models. Each panel corresponds to a predictive model and color represents the interpretable method (ALE-plot or PD-plot).}
    \label{fig-pdpalesup}
\end{figure}

Apartment area (\texttt{lsup\_constru}) is the most relevant feature in every model. The effect of this variable on the response is described with PD-plots and ALE-plots shown in Figure \ref{fig-pdpalesup} where each panel corresponds to a specific model. A negative effect is suggested by these plots with similar effect in every model. This is  true especially in the middle of the range for apartment area where there is most of the observed sample. Some differences can be seen in very large or very small apartments where random forest and XGboost show smaller effects on prices. %Estimated results for PD-plot and ALE-plot are very close. This might suggest the correlations among explanatory variables are not too high to bias the estimated effect with PD-plot and then the local correction made by ALE-plot has no effect. 

\begin{figure} 
    \centering
    \includegraphics[scale=0.5]{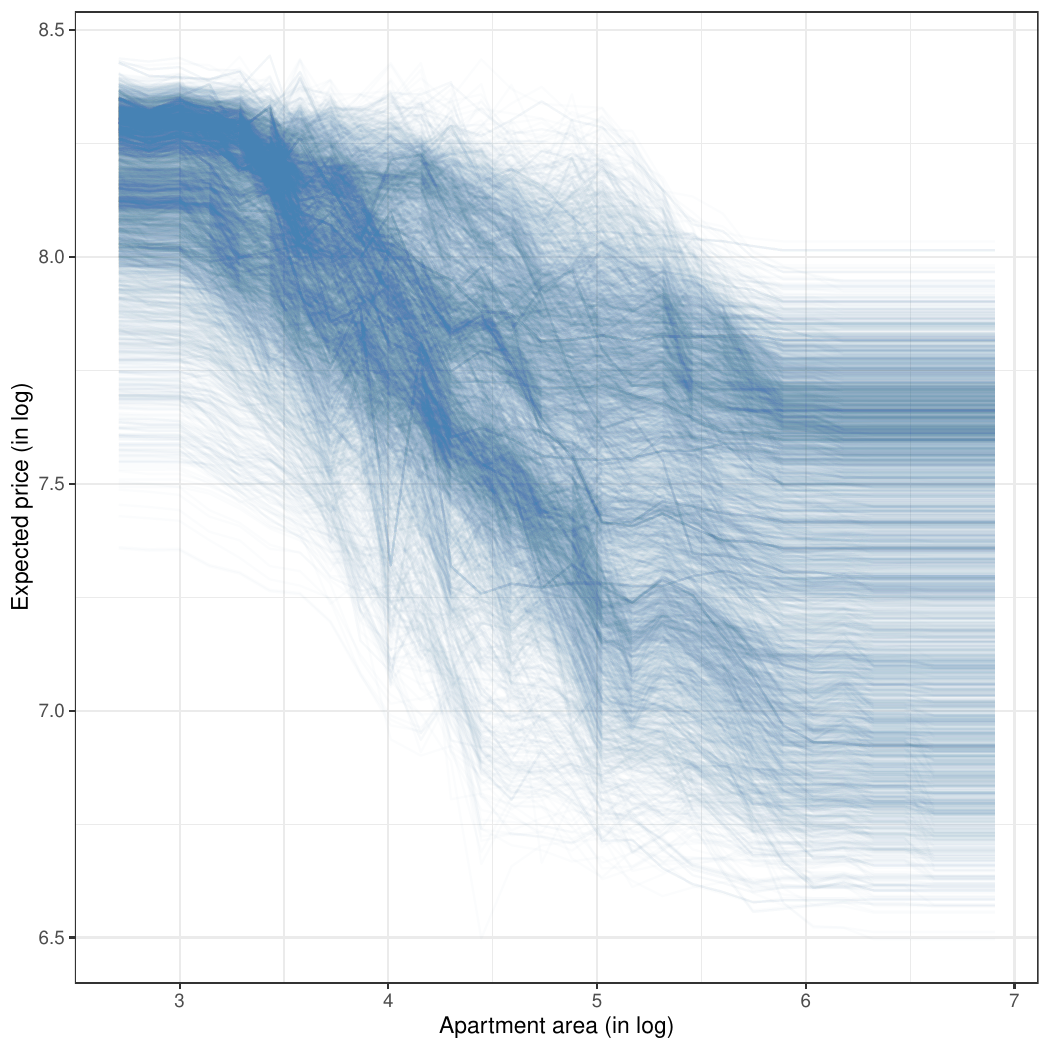}
    \caption{ICE-plot for \texttt{lsup\_constru} variable. Each panel corresponds to a predictive model. Showing 5000 randomly stratified selected curves.}
    \label{fig-icesupRF}
\end{figure}

Specifically, Figure \ref{fig-pdpalesup} shows the average effects of the variable \texttt{lsup\_constru}  in the sample. In a scenario where the impact of an explanatory variable over the response presents heterogeneity among observations, PD-plot and ALE-plot methods hide the variability of effects. An alternative method for interpretability to tackle this issue is the ICE-plot. The value of visualizing the individual curves that compose the PD-plot, is to explore other patterns in the effects than just the mean value.  Figure \ref{fig-icesupRF} shows the individual conditional expectation plot for  apartment area (\texttt{lsup\_constru}) for the  random forest model. The other predictive models shows similar results, as can be seen in Figure \ref{fig-icesup} in the Appendix. As it mentioned in Section \ref{section:2} there are too many ICE curves to display, using only transparency it was not enough to produce a good plot. ICE curves of a random sample using observed response deciles as strata are shown. 

The results illustrate the negative effect of the variable \texttt{lsup\_constru} in each individual curve. However, it is relevant to note that the effect  present  heterogeneity in the different properties. Figure \ref{fig-icesupRF} suggests that properties with higher predicted prices exhibit a small and linear effect, whereas cheaper properties show a non-linear relationship, with a larger impact of the apartment area.

\subsection{ SpICE curve effects } \label{results-ice}
Finally, to explore the connection among ICE curves and geographical location of properties, SpICE curves results are presented.  

In order to compute $D_1$ matrix (see Section \ref{seccD}), Euclidean distance between apartment coordinates is used. 
This metric is easy to visualize and applicable in the city of Montevideo due to the relatively close distances involved (maximum 30 km). However, it is possible to define other metrics using these two variables or other variables that better characterize the distance between two properties.

The range of the number of clusters is defined as three to five clusters. For each value within this range, an optimal value for the $\alpha$ parameter is determined as a compromise between the within-cluster homogeneity in terms of ICE curves and geographical proximity.
Results are shown in Figure \ref{alpfaoptimo} in the appendix, final choice is to work with four groups and $\alpha = 0.5$. 

\begin{figure} 
    \centering
    %{fig-4grupos.pdf}
    \includegraphics[scale=0.9]{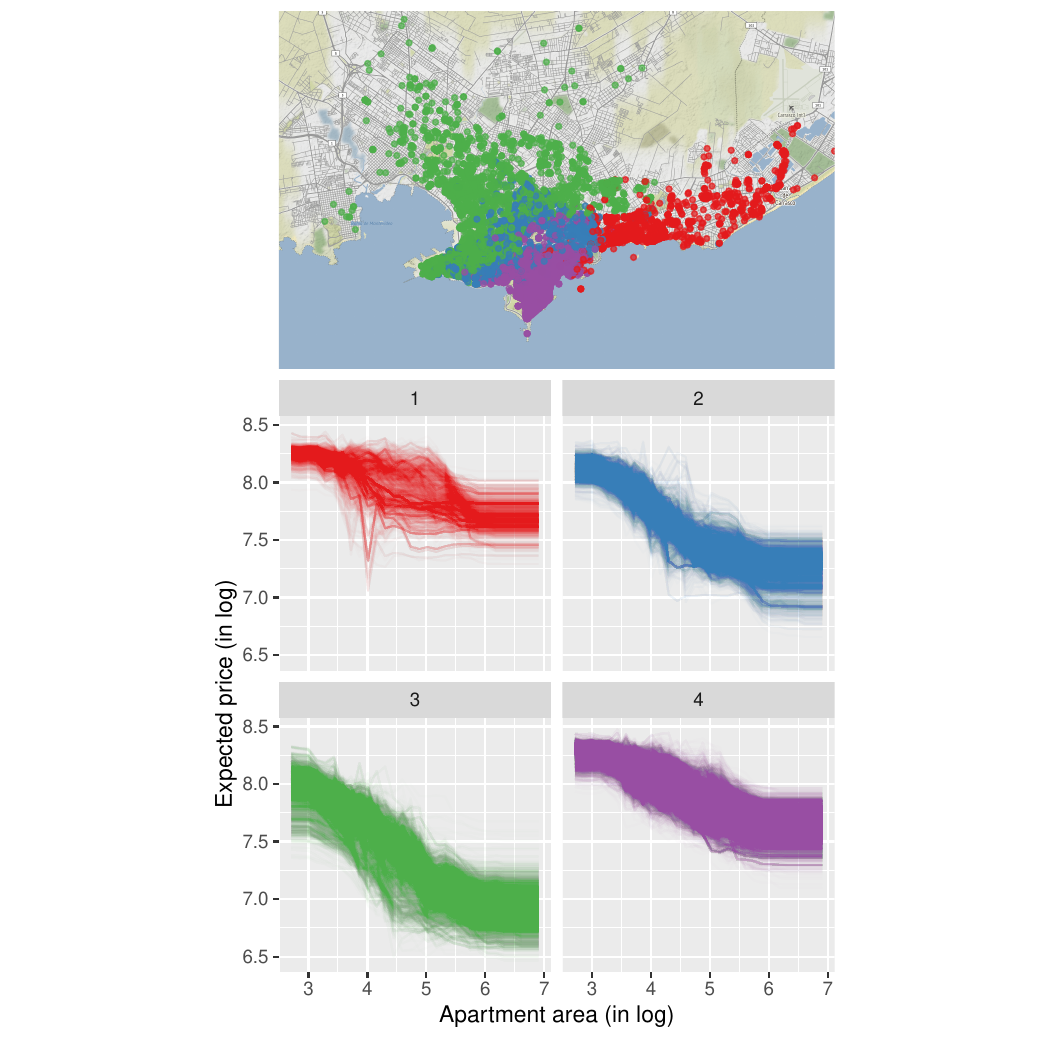}
    \caption{SpICE curves and geographical location of clusters.}
    \label{fig-spice}
\end{figure}

Figure \ref{fig-spice} shows the results of the geographically constrained clusters. Each apartment for sale in Montevideo city is represented as a point on the map, which is connected to an ICE curve in the bottom panel. The colors of the points and curves indicate the corresponding cluster assignments. Cluster location suggest a layout in the NorthWest-SouthEast direction, similar to the price gradient present in the data (see Figure \ref{mapa}). The apartments in the west and north-west side (green cluster) corresponds low income neighborhoods while the East side of the city (red cluster) represents the highest income zone in Montevideo.  

Focusing on the SpICE curves presented in Figure \ref{fig-spice}, there are distinct patterns in the relationship between apartment area and the asking price per square meter. Across all clusters, there is a negative relationship between the price per square meter and the apartment area. However, the effect of apartment area on square meter price differs between the green and blue clusters (associated with medium and low-income neighborhoods) and the red and violet clusters (associated with high-income neighborhoods). In green and blue clusters, an increase in apartment area results in a larger decrease in price per square meter compared to the red and violet clusters. This suggests that, in high-income neighborhoods, the total apartment price is more sensitive to changes in apartment area than in low-income neighborhoods where the total apartment prices are more inelastic to changes in apartment area. Consequently, the additional square meter in the apartment has no impact on the apartment price, leading to a decline in the square meter price within the low and medium-income neighborhoods.

% The properties maps show in the bottom left corner a few apartments located in a low income zone of the city, all points belong to the blue cluster but one which correspond to violet cluster. Figure \ref{fig-spice-zoom} shows SpICE curves for these apartments, as an example to illustrate the reason that geographically close point are classified into different groups. SpICE curve for the violet point show a linear negative effect of the total area, while blue curves of the close by apartments shows a highly non-linear effect of the apartment area. 
% \begin{figure} 
%     \centering
%     \includegraphics[scale=0.65]{fig-casa-cerro.pdf}
%     \caption{An example of ICE curves away (blue and violet) from spatially close apartments.}
%     \label{fig-spice-zoom}
% \end{figure}

\newpage

\section{Discussion} \label{section:4}
In this paper, model-agnostic interpretable methods for black box models were combined with spatial information to enhance interpretability in spatial applications. Specifically, the paper introduces the concept of geographically constrained clusters of ICE curves, referred to as SpICE curves.
Similar to ICE plots, SpICE curves can reveal heterogeneous effects of a predictor variable in a black box model.
The advantage of SpICE curves lies in their ability to interpret multiple clusters simultaneously, making them more suitable for large-scale data applications. Additionally, the spatial contiguity of the clusters provides additional interpretable information, further improving the analysis.

In the motivating example, SpICE curves were utilized to analyze the impact of the total area of properties on the price per square meter in apartments located in Montevideo city. Five statistical learning methods were chosen from a subset of 25 fitted models, based on their predictive performance using the \texttt{h2o} package with \texttt{autoML} procedure. The apartment's total area emerged as the most relevant feature in all five models, confirming its importance.

 Then using a fitted model with Random Forest algorithm, ICE-plot and SpICE curves for each apartment were computed. Property spatial information is combined with ICE curves to gain interpretability. Four property clusters were selected, based on a distance that combines a functional distance between ICE curves and Euclidean distance between apartment geographical coordinates.  Clusters blue and green mainly represent properties located in low or mid-low income zones in Montevideo and these clusters present a large, negative, non-linear effect of the apartment area on the price per square meter. On the other hand, clusters violet and red mainly correspond to mid high and high income neigborhoods, showing a small,  close to linear effect of the property area. 

This paper suggests several aspects that could be explored in future work. Firstly, the improvement of the functional distance used among ICE curves could be considered, such as the utilization of a weighted functional distance, as proposed in \cite{chen2014optimally}.  Secondly, instead of constructing fixed clusters of ICE curves, an alternative approach could involve considering local averages of the ICE curves in a nearest neighbors fashion,  where the distance used to determine the neighbors combines the structure of the ICE curves with geographical distance. Finally, a potential enhancement could involve summarizing the cluster ICE curves using ALE-plots, rather than relying on the average of clustered ICE curves. 

\section{Supplementary materials}
 This article was written with the R packages \texttt{knitr} \citep{knitr}, \texttt{ggplot2} \citep{hadley:2016}, \texttt{leaflet} \citep{leaflet}, \texttt{tidyverse} \citep{tidyverse}, \texttt{h2o} \citep{h2o}, \texttt{ClustGeo} \citep{clustGeo},
 \texttt{KernSmooth} \citep{KernSmooth} and the files to reproduce the article and results is available at https://github.com/natydasilva/SpICE\_COST. 

\renewcommand{\refname}{References }
\pagebreak
\bibliographystyle{chicago}
\bibliography{bibcsic.bib}

\begin{thebibliography}{}

\bibitem[\protect\citeauthoryear{Adams and Fournier}{Adams and Fournier}{2003}]{sobolev}
Adams, R.~A. and J.~J.~F. Fournier (2003).
\newblock {\em Sobolev spaces}.
\newblock Elsevier/Academic Press.

\bibitem[\protect\citeauthoryear{Apley}{Apley}{2018}]{aleplot}
Apley, D. (2018).
\newblock {\em ALEPlot: Accumulated Local Effects (ALE) Plots and Partial Dependence (PD) Plots}.
\newblock R package version 1.1.

\bibitem[\protect\citeauthoryear{Apley and Zhu}{Apley and Zhu}{2020}]{apley2020visualizing}
Apley, D.~W. and J.~Zhu (2020).
\newblock Visualizing the effects of predictor variables in black box supervised learning models.
\newblock {\em Journal of the Royal Statistical Society: Series B (Statistical Methodology)\/}~{\em 82\/}(4), 1059--1086.

\bibitem[\protect\citeauthoryear{Case, Clapp, Dubin, and Rodriguez}{Case et~al.}{2004}]{case2004}
Case, B., J.~Clapp, R.~Dubin, and M.~Rodriguez (2004).
\newblock Modeling spatial and temporal house price patterns: A comparison of four models.
\newblock {\em The Journal of Real Estate Finance and Economics\/}~{\em 29\/}(2), 167--191.

\bibitem[\protect\citeauthoryear{Chavent, Kuentz, Labenne, and Saracco}{Chavent et~al.}{2021}]{clustGeo}
Chavent, M., V.~Kuentz, A.~Labenne, and J.~Saracco (2021).
\newblock {\em ClustGeo: Hierarchical Clustering with Spatial Constraints}.
\newblock R package version 2.1.

\bibitem[\protect\citeauthoryear{Chavent, Kuentz-Simonet, Labenne, and Saracco}{Chavent et~al.}{2018}]{chavent2018}
Chavent, M., V.~Kuentz-Simonet, A.~Labenne, and J.~Saracco (2018).
\newblock Clustgeo: an r package for hierarchical clustering with spatial constraints.
\newblock {\em Computational Statistics\/}~{\em 33\/}(4), 1799--1822.

\bibitem[\protect\citeauthoryear{Chen, Reiss, and Tarpey}{Chen et~al.}{2014}]{chen2014optimally}
Chen, H., P.~T. Reiss, and T.~Tarpey (2014).
\newblock Optimally weighted l2 distance for functional data.
\newblock {\em Biometrics\/}~{\em 70\/}(3), 516--525.

\bibitem[\protect\citeauthoryear{Cheng, Karambelkar, and Xie}{Cheng et~al.}{2022}]{leaflet}
Cheng, J., B.~Karambelkar, and Y.~Xie (2022).
\newblock {\em leaflet: Create Interactive Web Maps with the JavaScript 'Leaflet' Library}.
\newblock R package version 2.1.1.

\bibitem[\protect\citeauthoryear{Dubin}{Dubin}{1988}]{dubin1988estimation}
Dubin, R.~A. (1988).
\newblock Estimation of regression coefficients in the presence of spatially autocorrelated error terms.
\newblock {\em The Review of Economics and Statistics\/}, 466--474.

\bibitem[\protect\citeauthoryear{Friedman}{Friedman}{2001}]{friedman2001}
Friedman, J.~H. (2001).
\newblock Greedy function approximation: a gradient boosting machine.
\newblock {\em Annals of statistics\/}, 1189--1232.

\bibitem[\protect\citeauthoryear{Goldstein, Kapelner, Bleich, and Pitkin}{Goldstein et~al.}{2015}]{goldstein2015peeking}
Goldstein, A., A.~Kapelner, J.~Bleich, and E.~Pitkin (2015).
\newblock Peeking inside the black box: Visualizing statistical learning with plots of individual conditional expectation.
\newblock {\em Journal of Computational and Graphical Statistics\/}~{\em 24\/}(1), 44--65.

\bibitem[\protect\citeauthoryear{Goyeneche, Moreno, and Scavino}{Goyeneche et~al.}{2017}]{goyeneche2017}
Goyeneche, J.~J., L.~Moreno, and M.~Scavino (2017).
\newblock Predicci{\'o}n del valor de un inmueble mediante t{\'e}cnicas agregativas.
\newblock {\em Serie DT (17/1)\/}.

\bibitem[\protect\citeauthoryear{Greenwell}{Greenwell}{2017}]{pdp}
Greenwell, B.~M. (2017).
\newblock pdp: An {R} package for constructing partial dependence plots.
\newblock {\em The R Journal\/}~{\em 9\/}(1), 421--436.

\bibitem[\protect\citeauthoryear{Grubesic, Wei, and Murray}{Grubesic et~al.}{2014}]{grubesic2014}
Grubesic, T.~H., R.~Wei, and A.~T. Murray (2014).
\newblock Spatial clustering overview and comparison: Accuracy, sensitivity, and computational expense.
\newblock {\em Annals of the Association of American Geographers\/}~{\em 104\/}(6), 1134--1156.

\bibitem[\protect\citeauthoryear{Hall, Gill, Kurka, and Phan}{Hall et~al.}{2017}]{hall2017}
Hall, P., N.~Gill, M.~Kurka, and W.~Phan (2017).
\newblock Machine learning interpretability with h2o driverless ai.
\newblock {\em H2O. ai\/}.

\bibitem[\protect\citeauthoryear{Kiel and Zabel}{Kiel and Zabel}{2008}]{kiel2008}
Kiel, K.~A. and J.~E. Zabel (2008).
\newblock Location, location, location: The 3l approach to house price determination.
\newblock {\em Journal of Housing Economics\/}~{\em 17\/}(2), 175--190.

\bibitem[\protect\citeauthoryear{Kuhn}{Kuhn}{2021}]{caret}
Kuhn, M. (2021).
\newblock {\em caret: Classification and Regression Training}.
\newblock R package version 6.0-90.

\bibitem[\protect\citeauthoryear{Kuhn and Wickham}{Kuhn and Wickham}{2020}]{tidymodels}
Kuhn, M. and H.~Wickham (2020).
\newblock {\em Tidymodels: a collection of packages for modeling and machine learning using tidyverse principles.}

\bibitem[\protect\citeauthoryear{LeDell, Gill, Aiello, Fu, Candel, Click, Kraljevic, Nykodym, Aboyoun, Kurka, and Malohlava}{LeDell et~al.}{2023}]{h2o}
LeDell, E., N.~Gill, S.~Aiello, A.~Fu, A.~Candel, C.~Click, T.~Kraljevic, T.~Nykodym, P.~Aboyoun, M.~Kurka, and M.~Malohlava (2023).
\newblock h2o: R interface for the 'h2o' scalable machine learning platform.
\newblock R package version 3.40.0.1.

\bibitem[\protect\citeauthoryear{Limsombunchai}{Limsombunchai}{2004}]{limsombunchai2004}
Limsombunchai, V. (2004).
\newblock House price prediction: hedonic price model vs. artificial neural network.
\newblock In {\em New Zealand agricultural and resource economics society conference}, pp.\  25--26.

\bibitem[\protect\citeauthoryear{Maksymiuk, Gosiewska, and Biecek}{Maksymiuk et~al.}{2020}]{maksymiuk2020landscape}
Maksymiuk, S., A.~Gosiewska, and P.~Biecek (2020).
\newblock Landscape of r packages for explainable artificial intelligence.
\newblock {\em arXiv preprint arXiv:2009.13248\/}.

\bibitem[\protect\citeauthoryear{Miller}{Miller}{2019}]{miller2019explanation}
Miller, T. (2019).
\newblock Explanation in artificial intelligence: Insights from the social sciences.
\newblock {\em Artificial intelligence\/}~{\em 267}, 1--38.

\bibitem[\protect\citeauthoryear{Molnar, Casalicchio, and Bischl}{Molnar et~al.}{2020}]{molnar2020}
Molnar, C., G.~Casalicchio, and B.~Bischl (2020).
\newblock Interpretable machine learning--a brief history, state-of-the-art and challenges.
\newblock In {\em Joint European Conference on Machine Learning and Knowledge Discovery in Databases}, pp.\  417--431. Springer.

\bibitem[\protect\citeauthoryear{Mooya}{Mooya}{2016}]{mooya2016standard}
Mooya, M.~M. (2016).
\newblock Standard theory of real estate market value: Concepts and problems.
\newblock {\em Real Estate Valuation Theory: A Critical Appraisal\/}, 1--21.

\bibitem[\protect\citeauthoryear{Osland}{Osland}{2010}]{osland2010}
Osland, L. (2010).
\newblock An application of spatial econometrics in relation to hedonic house price modeling.
\newblock {\em Journal of Real Estate Research\/}~{\em 32\/}(3), 289--320.

\bibitem[\protect\citeauthoryear{Park and Bae}{Park and Bae}{2015}]{park2015}
Park, B. and J.~K. Bae (2015).
\newblock Using machine learning algorithms for housing price prediction: The case of fairfax county, virginia housing data.
\newblock {\em Expert systems with applications\/}~{\em 42\/}(6), 2928--2934.

\bibitem[\protect\citeauthoryear{Picardo}{Picardo}{2019}]{Picardo}
Picardo, P. (2019).
\newblock Predicción de precios de la vivienda aprendizaje estadístico con datos de ofertas y transacciones para montevideo.
\newblock {\em Tesis de Maestría en Economía, FCEA-UDELAR\/}.

\bibitem[\protect\citeauthoryear{Rosen}{Rosen}{1974}]{rosen1974}
Rosen, S. (1974).
\newblock Hedonic prices and implicit markets: Product differentiation in pure competition.
\newblock {\em Journal of Political Economy\/}~{\em 82\/}(1), 34--55.

\bibitem[\protect\citeauthoryear{Sirmans, Macpherson, and Zietz}{Sirmans et~al.}{2005}]{sirmans2005}
Sirmans, S., D.~Macpherson, and E.~Zietz (2005).
\newblock The composition of hedonic pricing models.
\newblock {\em Journal of real estate literature\/}~{\em 13\/}(1), 1--44.

\bibitem[\protect\citeauthoryear{Varghese, Unnikrishnan, and Jacob}{Varghese et~al.}{2013}]{varghese2013}
Varghese, B.~M., A.~Unnikrishnan, and K.~Jacob (2013).
\newblock Spatial clustering algorithms-an overview.
\newblock {\em Asian Journal of Computer Science and Information Technology\/}~{\em 3\/}(1), 1--8.

\bibitem[\protect\citeauthoryear{Wand}{Wand}{2021}]{KernSmooth}
Wand, M. (2021).
\newblock {\em KernSmooth: Functions for Kernel Smoothing Supporting Wand \& Jones (1995)}.
\newblock R package version 2.23-20.

\bibitem[\protect\citeauthoryear{Wickham}{Wickham}{2016}]{hadley:2016}
Wickham, H. (2016).
\newblock {\em ggplot2: Elegant Graphics for Data Analysis}.
\newblock Springer-Verlag New York.

\bibitem[\protect\citeauthoryear{Wickham, Averick, Bryan, Chang, McGowan, François, Grolemund, Hayes, Henry, Hester, Kuhn, Pedersen, Miller, Bache, Müller, Ooms, Robinson, Seidel, Spinu, Takahashi, Vaughan, Wilke, Woo, and Yutani}{Wickham et~al.}{2019}]{tidyverse}
Wickham, H., M.~Averick, J.~Bryan, W.~Chang, L.~D. McGowan, R.~François, G.~Grolemund, A.~Hayes, L.~Henry, J.~Hester, M.~Kuhn, T.~L. Pedersen, E.~Miller, S.~M. Bache, K.~Müller, J.~Ooms, D.~Robinson, D.~P. Seidel, V.~Spinu, K.~Takahashi, D.~Vaughan, C.~Wilke, K.~Woo, and H.~Yutani (2019).
\newblock Welcome to the {tidyverse}.
\newblock {\em Journal of Open Source Software\/}~{\em 4\/}(43), 1686.

\bibitem[\protect\citeauthoryear{Xie}{Xie}{2023}]{knitr}
Xie, Y. (2023).
\newblock {\em knitr: A General-Purpose Package for Dynamic Report Generation in R}.
\newblock R package version 1.42.

\bibitem[\protect\citeauthoryear{Yoo, Im, and Wagner}{Yoo et~al.}{2012}]{yoo2012}
Yoo, S., J.~Im, and J.~E. Wagner (2012).
\newblock Variable selection for hedonic model using machine learning approaches: A case study in onondaga county, ny.
\newblock {\em Landscape and Urban Planning\/}~{\em 107\/}(3), 293--306.

\end{thebibliography}

\appendix
\section{Supplementary figures}

\begin{figure} 
    \centering
    \includegraphics[scale=0.65]{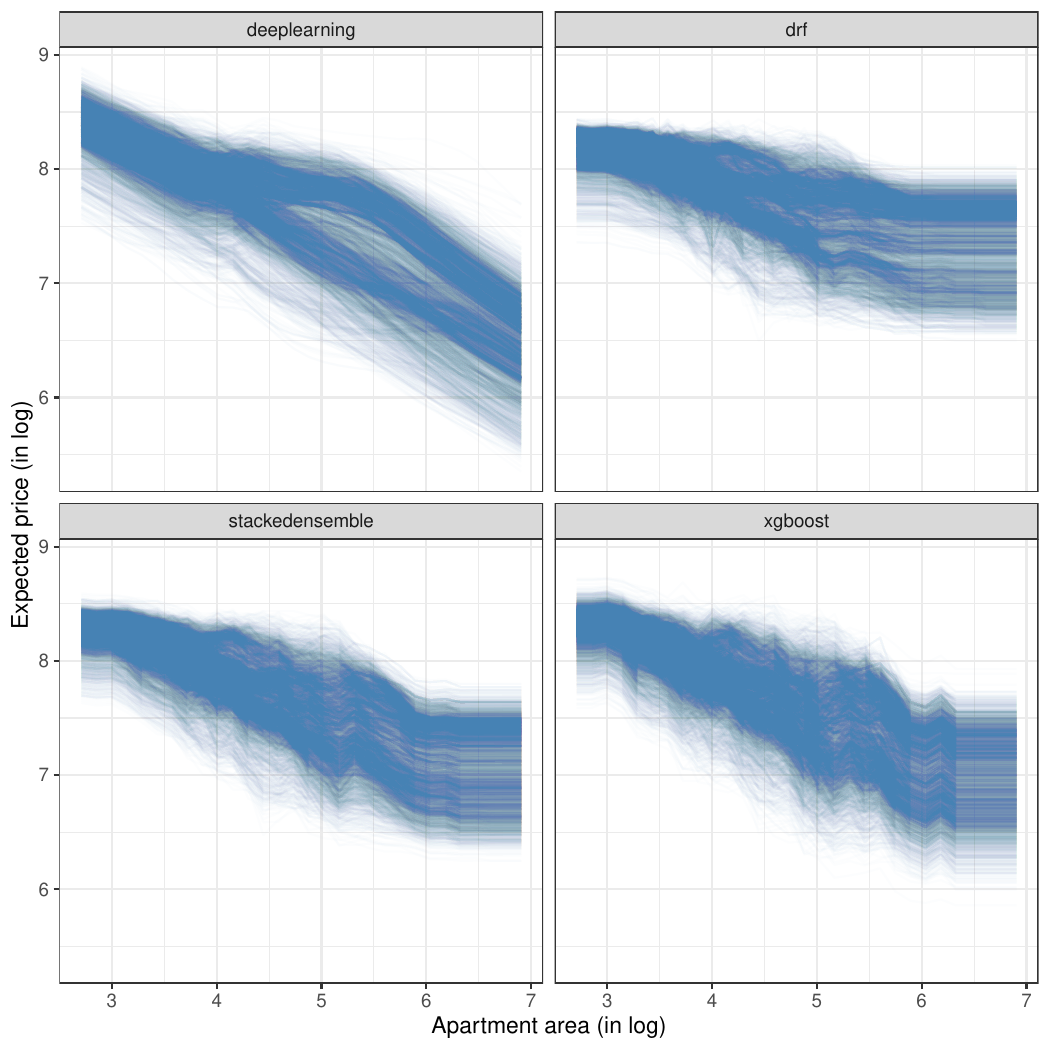}
    \caption{ICE-plot for \textit{log apartment area} variable. Each panel corresponds to a predictive model. Showing 5000 randomly stratified selected curves.}
    \label{fig-icesup}
\end{figure}

\begin{figure} 
    \centering
    \includegraphics[scale=.6]{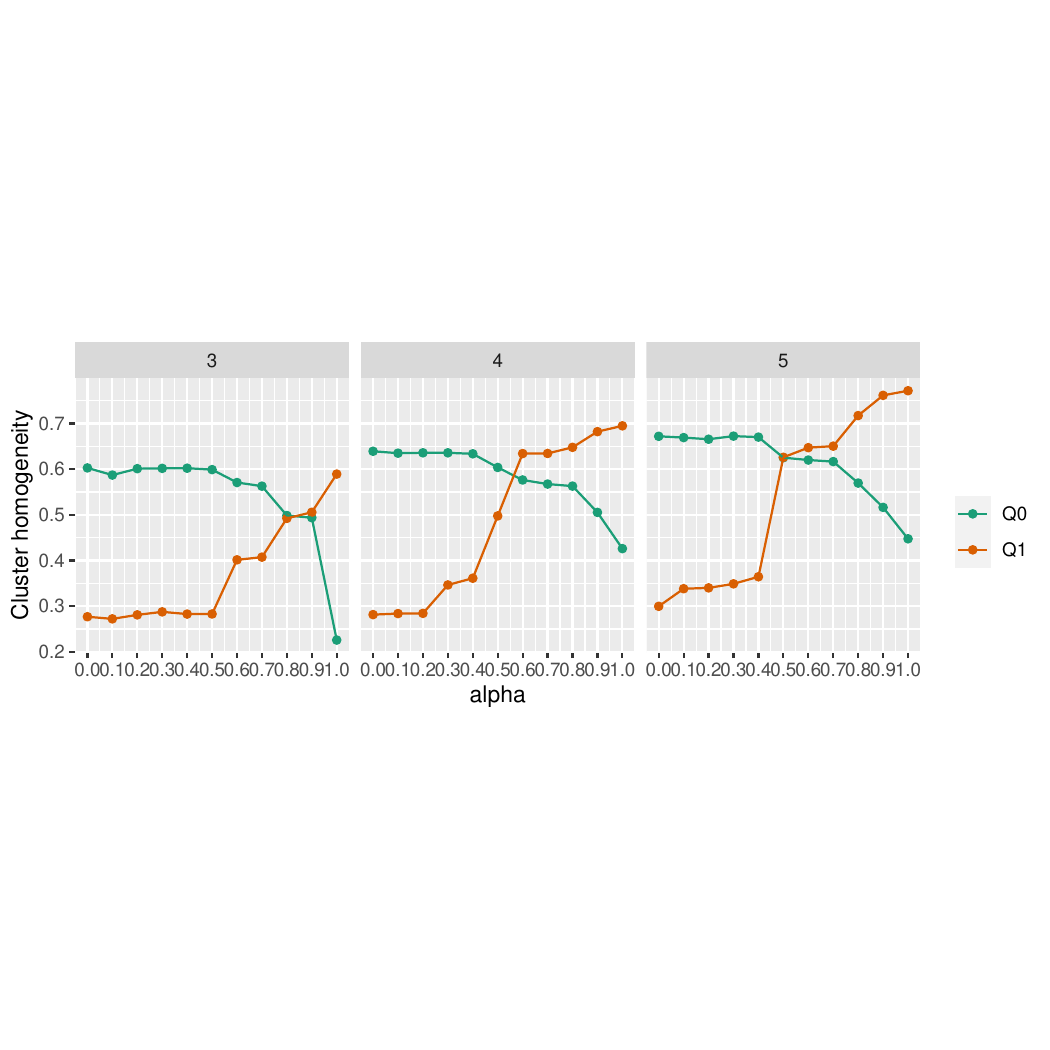}
    \caption{ Optimal $\alpha$ for different groups. }
    \label{alpfaoptimo}
\end{figure}

\end{document}